\documentclass[intlimits,twoside,a4paper]{article}

\usepackage[cp1251]{inputenc}

\usepackage{cmpj3}

\usepackage{bm}

\issue{2019}{22}{4}{43605}
\doinumber{10.5488/CMP.22.43605}

\title[Caged dynamics and JGX $\beta$-relaxation in glass transition problem]%
{Absolutely necessary to consider the caged dynamics and the JGX $\beta$-relaxation in solving the glass transition problem}
\author[K.L. Ngai, S. Capaccioli, L.M. Wang]%
{K.L. Ngai\refaddr{label1,label2}, S. Capaccioli\refaddr{label3}, L.M. Wang\refaddr{label2}}
\addresses{%
\addr{label1} CNR-IPCF, Largo Bruno Pontecorvo 3, I-56127 Pisa, Italy%
\addr{label2} State Key Lab of Metastable Materials Science and Technology, Yanshan 
University, Qinhuangdao, Hebei 066004, China%
\addr{label3} Dipartimento di Fisica, Universit{\`a} di Pisa, Largo Bruno Pontecorvo 3, I-56127 Pisa, Italy%
}

\date{Received June 3, 2019}

\begin{document}

\maketitle

\begin{abstract}
The 2003 seminal paper entitled ``Is the Fragility of a Liquid Embedded in the Properties of Its Glass?'' by Tullio Scopigno, Giancarlo Ruocco, Francesco Sette, and Giulio Monaco, reported that the properties of the structural $\alpha$-relaxation of glass-formers are already present in the faster caged dynamics. Their important discovery has far-reaching implication of the processes faster than the structural $\alpha$-relaxation that cannot be ignored in solving the glass transition problem. Since then, experiments and simulations performed on many glass-formers with diverse chemical and physical structures have found strong connections of the $\alpha$-relaxation with not only the caged dynamics but also with the secondary relaxation of the special kind, called the JGX $\beta$-relaxation. The general results indicate that these fast processes are inseparable from the $\alpha$-relaxation, and any attempt to solve the glass transition problem should take  account of this fact. Examples of the connections are given in this paper to elucidate the developments and advances made since the inspiring publication of Scopigno et al.
\pacs 64.70.Pf, 77.22.Gm, 78.70.Ck, 78.70.Nx, 71.23.Cq
\end{abstract}

\section{Introduction}

Published in 2003, the paper ``Is the Fragility of a Liquid Embedded in the Properties of Its Glass?'' by Tullio Scopigno, Giancarlo Ruocco, Francesco Sette, and Giulio Monaco \cite{NCW_bib01} is a landmark publication with impact on past and current research in dynamics and thermodynamics of glass-forming materials. In this innovative research they used inelastic x-ray scattering (IXS) measurements of the dynamic structure factor to access the high frequency (terahertz) dynamical properties of glasses of different kinds. The IXS measurements enabled the determination of the nonergodicity factor $f(Q,T)$, with a reliability not achievable by light or neutron scattering \cite{NCW_bib02}. They found that the IXS data of $f(Q,T)$ in the low $Q$ limit at temperatures below the glass transition temperature $T_{\text{g}}$ of many glass-formers can be well described by the functional form,
\begin{equation}
	f(Q\to 0,T)=[1+\alpha(T/T_{\text{g}})]^{-1},
\label{NCW_eqn01}
\end{equation}
where $\alpha$ is a material dependent parameter. From the collection of data plotted as $1/f(Q^*,T)$ vs. $T/T_{\text{g}}$ with for $Q^*=2$~nm$^{-1}$, they found the existence of a strong correlation between $\alpha$ and $m$, the fragility index \cite{NCW_bib03}. Thus, they found that the vibrational properties of the glass well below $T_{\text{g}}$ correlate with the fragility index, which is a property of the cooperative $\alpha$-relaxation of the viscous liquid at and above~$T_{\text{g}}$.

These seminal research results from Scopigno, Ruocco, Sette, and Monaco (SRSM) imply that faster processes that transpire before the many-molecules cooperative $\alpha$-relaxation cannot be ignored if the goal is a complete and fundamental account of the glass transition problem. In this paper, we consider other experimental findings and molecular dynamic simulations that are consistent with the work by SRSM. More importantly, we present the other connections of the short time vibrational and secondary relaxation to the terminal $\alpha$-relaxation developed over two decades, which magnify the impact.

\section{Supporting IXS data of SRSM by neutron scattering data and simulations}

\begin{figure}[!b]
\centerline{\includegraphics[width=0.5\textwidth]{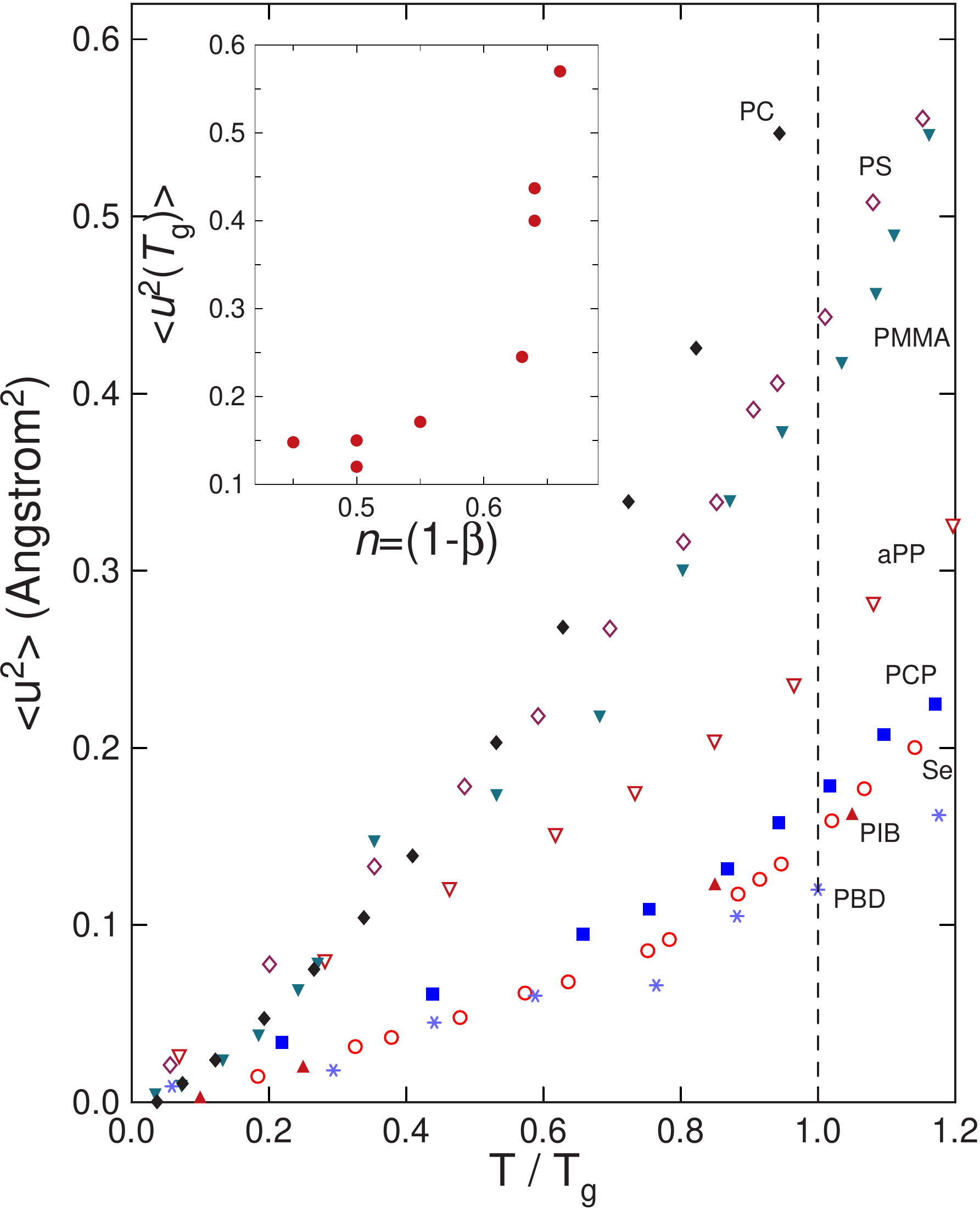}}
\caption{(Colour online) The mean squared displacement $\langle u^{2}(T)\rangle$ as a function of $T/T_{\text{g}}$ of several amorphous polymers, PC, PMMA, 1,4-PB, and Se obtained from incoherent neutron scattering by the IN6 spectrometer of the Institut Laue-Langevin together with the data of PCP, PS, aPP, and PIB obtained by Kanaya and co-worker with spectrometers having about the same energy resolution as IN6 (0.2~meV)~\cite{NCW_bib06}. The sources of the experimental data can be found in reference~\protect\cite{NCW_bib06}. The inset is a plot of $\langle u^{2}(T_{\text{g}})\rangle$ vs.~$n$.}
\label{NCW_fig01}
\end{figure}

The nonergodicity factor $f(Q,T)$ is the long time limit of the density-density correlation function $F(Q,t)$, normalized to the static structure factor $S(Q)$ \cite{NCW_bib01,NCW_bib04}. In neutron scattering at wavevector $Q$ \cite{NCW_bib05}, the non-ergodicity parameter or the effective Debye-Waller factor is observed from the $T$-dependence of the mean-square-displacement (MSD), $\langle u^{2}(T)\rangle$, by the expression
\begin{equation}
	f(Q,T)=\exp[-\langle u^{2}(T)\rangle Q^2/3].
\label{NCW_eqn02}
\end{equation}
Hence, we can see that the correlation between $\alpha$ and $m$ is closely related to another correlation of either $m$ or $n$ with $\langle u^{2}(T/T_{\text{g}})\rangle$ found by comparing data of many glass-formers from quasielastic neutron scattering measurements at temperatures below $T_{\text{g}}$ \cite{NCW_bib06}. Here, $n$ is the complement of the fractional exponent $(1-n)\equiv \beta_{\text{K}}$ in the Kohlrausch correlation function of the structural $\alpha$-relaxation
\begin{equation}
	\phi(t) = \exp[-(t/\tau_{\alpha})^{1-n}].
\label{NCW_eqn03}
\end{equation}
That is, the glass former with larger $m$ or $n$ has a larger $\langle u^{2}(T)\rangle$ at the same value of $T/T_{\text{g}}$ and rises more rapidly as a function of $T/T_{\text{g}}$, below $T_{\text{g}}$ as well as near and across $T_{\text{g}}$ in the liquid states \cite{NCW_bib06}. An example of the correlation is shown in figure~\ref{NCW_fig01} for polymers, PIB (0.45), cis 1,4-PB (0.48), Se (0.50), trans-1,4-polychloroprene (PCP) (0.55), atatic polypropylene (aPP) (0.63), PMMA (0.63), PS (0.64), and bisphenol A polycarbonate (PC) (0.65). Inside the brackets there are the $n$ values of the corresponding polymer. By inspection of figure~\ref{NCW_fig01}, one can verify, for any scaled temperature $T/T_{\text{g}}$ that $\langle u^{2}(T/T_{\text{g}})\rangle$, correlated approximated with $n$, the non-exponentiality parameter of the $\alpha$-relaxation. The inset shows the correlation of $\langle u^{2}(T/T_{\text{g}})\rangle$ at $T/T_{\text{g}}=1$ with $n$. All $\langle u^{2}(T/T_{\text{g}})\rangle$ data shown were taken by the spectrometers with energy resolution of 0.2 meV to ensure an objective comparison, since neutron data depend on the energy resolution.

Note  that $\langle u^{2}(T)\rangle$ changes to a stronger temperature dependence in the liquid state after crossing $T_{\text{g}}$ from below. As will be discussed in the next section, the $\langle u^{2}(T)\rangle$ measured near $T_{\text{g}}$ by neutron scattering at time scales usually shorter than 1~ns correspond to the motions of molecules within cages affected by the intermolecular potential. Thus, the change of $\langle u^{2}(T)\rangle$ at $T_{\text{g}}$ defined by the $\alpha$-relaxation means that the caged molecule dynamics is sensitive to the density change, a point which will be revisited in the sections follow.

Ultimately, the correlations found must come from the intermolecular potential and its variation from one glass-former to another, a point also recognized by SRSM. This prompted the performance of molecular dynamics simulations of binary Lennard-Jones systems with three different potentials with increasing anharmonicity \cite{NCW_bib07,NCW_bib08}. The results show that the increase of anharmonicity and capacity for intermolecular coupling of the potential is the cause of the increase of kinetic fragility and non-exponentiality in the liquid state, and in parallel the $T_{\text{g}}$-scaled temperature dependence of $1/f(Q,T)$ vs.\ $T/T_{\text{g}}$ determined by the vibrations at low temperatures in the glassy state. Hence, the basis of the correlation discovered experimentally by SRSM is on anharmonicity of the intermolecular potential on which the coupling model (CM) was built \cite{NCW_bib09,NCW_bib10,NCW_bib11,NCW_bib12,NCW_bib13}.

\section{The JGX $\beta$-relaxation: the intermediary of caged dynamics and $\alpha$-relaxation}
\label{NCW_sctn03}

The correlations found by Scopigno et al. \cite{NCW_bib01} and by neutron scattering \cite{NCW_bib06} imply that the $\alpha$-relaxation is linked to the faster process manifested by $1/f(Q\to 0,T)$ or $\langle u^{2}(T)\rangle$. These latter quantities in the glass state obtained by the IXS and neutron scattering solely pertain to the dynamics of molecules mutually caged via the intermolecular potential because the secondary relaxation time $\tau_{\beta}(T)$ and, even more, the primary relaxation time $\tau_{\alpha}(T)$ are much longer than the time-scales of measurements. The dynamics of caged molecules is not a relaxation or diffusion process and it has no characteristic times. Observed in susceptibility \cite{NCW_bib14}, the caged dynamics has the frequency dependence,
$\chi''(\omega)=A(T)\omega^{-c}$,
where $0<c\ll 1$ is a small positive constant, and the behaviour is appropriately called the nearly constant loss (NCL). It was also found by dielectric relaxation \cite{NCW_bib15,NCW_bib16,NCW_bib17}, light-scattering \cite{NCW_bib18,NCW_bib19}, and in optically heterodyne detected optical Kerr effect (OHD-OKE) experiments \cite{NCW_bib20,NCW_bib21,NCW_bib22}.

Notwithstanding the significance of the short time caged molecule dynamics, the question that arises quite naturally is whether a secondary relaxation exists in between it and the $\alpha$-relaxation which has properties that are correlated with those of the former as well as the latter? If such a secondary relaxation exists, it would bridge the link of the caged dynamics to the $\alpha$-relaxation, and make the physics richer. A development by the CM starting in 1998 \cite{NCW_bib13,NCW_bib23,NCW_bib24,NCW_bib25} has indeed found such secondary relaxations with properties strongly related to the $\alpha$-relaxation. Such secondary relaxations have relaxation times $\tau_{\beta}$ approximately the same as the primitive relaxation time of the CM, i.e.,
\begin{equation}
	\tau_{\beta} \approx \tau_{0}\,,
\label{NCW_eqn04}
\end{equation}
and it shifts on elevating pressure in concert with $\tau_{\alpha}$. To distinguish the secondary relaxations belonging to this special class from other trivial ones, they were called the Johari-Goldstein (JG) $\beta$-relaxations with the intention of honouring these two colleagues for their contributions to secondary relaxations and particularly to the discovery of a secondary relaxation in totally rigid molecule without internal degree of freedom. In retrospect, this is not a good choice because many researchers use the term, JG relaxation, for all secondary relaxations without distinction. A better choice of nomenclature for secondary relaxation in the special class is JGX $\beta$-relaxation, which we shall adopt from now on.

\begin{figure}[!b]
\centerline{\includegraphics[width=0.99\textwidth]{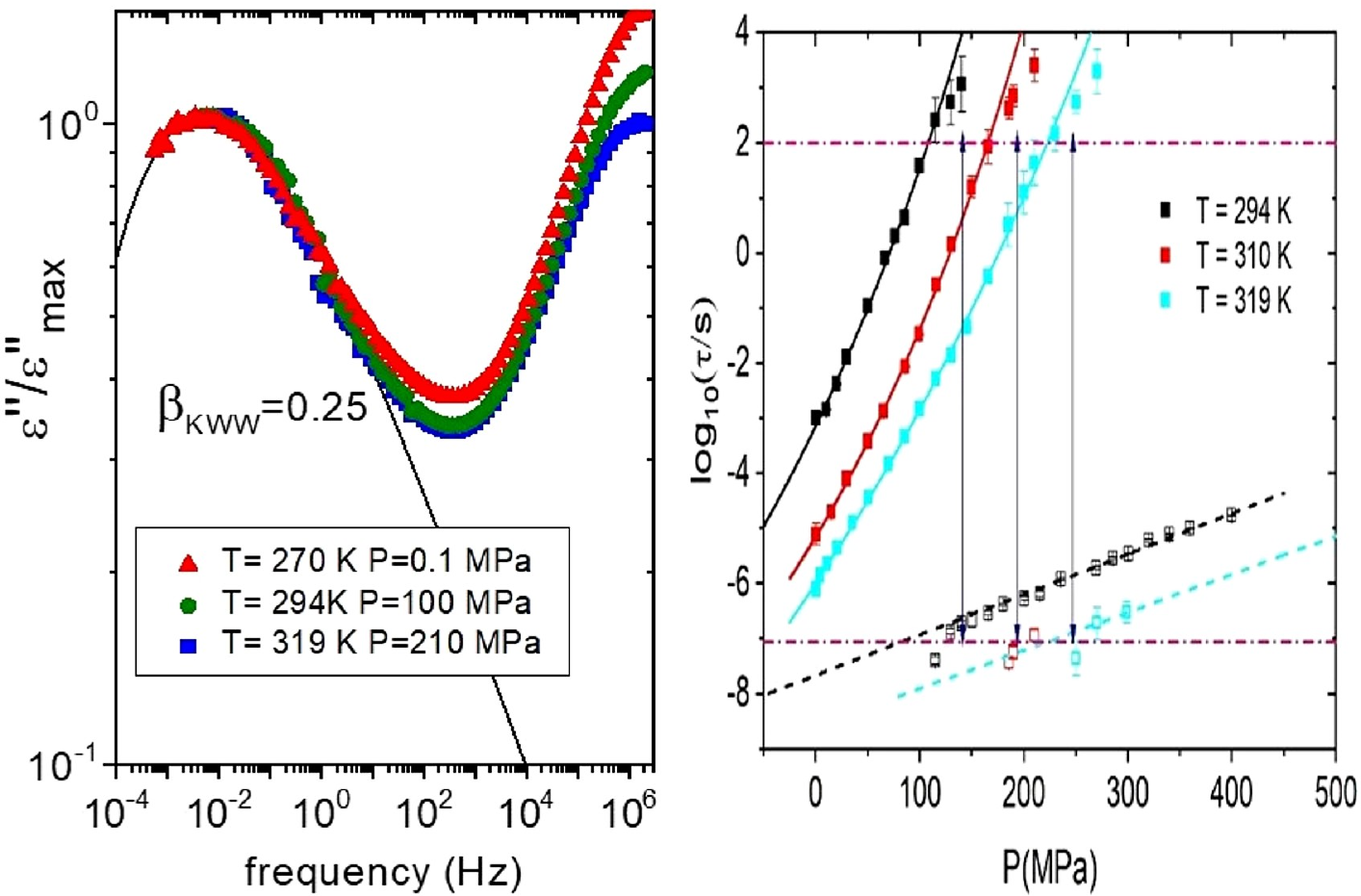}}
\caption{(Colour online) (Left-hand panel) Superposition of spectra at different pressures and temperatures, for the same frequency of the maximum of the peak, for the 20{\%}~TPP in PS mixture. The conductivity has been subtracted. Solid black lines are fits by the Fourier transform of the KWW function. (Right-hand panel) Relaxation map for 20{\%}~TPP in PS mixture as function of pressure at three different temperatures. Solid points represent relaxation times of the $\alpha$-relaxation, while open points represent relaxation times of the JG $\beta$-relaxation. Also shown are the fits of the $\alpha$-relaxation with the Vogel-Fulcher-Tammann equation (solid lines), while the JG $\beta$-relaxation times have been fitted with an Arrhenius law.}
\label{NCW_fig02}
\end{figure}

The primitive relaxation time $\tau_{0}(T,P)$ of the CM bears a rigorous connection to $\tau_{\alpha}(T,P)$ in temperature and pressure dependences via the time honoured CM equation \cite{NCW_bib09,NCW_bib13},
\begin{equation}
	\tau_{\alpha}(T,P)
	=
	\left[t_{c}^{-n}\tau_{0}(T,P)\right]^{1/(1-n)}.
\label{NCW_eqn05}
\end{equation}
In this equation, $t_{c} = 2$~ps and the Kohlrausch exponent $(1-n)$ or the frequency dispersion of the $\alpha$-relaxation is constant to variations of $T$ and $P$ at constant $\tau_{\alpha}(T,P)$ \cite{NCW_bib26}. Hence, by combining equations~\eqref{NCW_eqn04} and \eqref{NCW_eqn05}, we arrive at the prediction of approximate invariance of the separation between the $\alpha$-relaxation from the primitive relaxation given by
\begin{equation}
	\log[\tau_{\alpha}(P,T)]-\log[\tau_{\beta}(P,T)]
	\approx 
	n\left\{\log[\tau_{\alpha}(P,T)]-11.7\right\}
\label{NCW_eqn06}
\end{equation}
to variations of $T$ and $P$ as long as $\tau_{\alpha}(T,P)$ is kept constant. This prediction had been verified by experiments multiple times \cite{NCW_bib27,NCW_bib28,NCW_bib29,NCW_bib30,NCW_bib31,NCW_bib32,NCW_bib33,NCW_bib34}. Shown as an example, it is taken from the dielectric relaxation measurements at elevated pressures on a mixture of 20{\%} TPP (tripropyl phosphate) with polystyrene (PS) \cite{NCW_bib34}. The dielectric spectra in figure~\ref{NCW_fig02} show the faster JGX $\beta$-relaxation and the slower $\alpha$-relaxations at three different combinations of $T$ and $P$ while keeping the $\alpha$-loss peak frequency $f_{\alpha}(T,P)$ constant. The constancy of the corresponding $f_{\beta}(T,P)$ of the JGX $\beta$-relaxation verifies equation~\eqref{NCW_eqn06}. The right-hand panel of figure~\ref{NCW_fig02} shows the relaxation times, $\tau_{\alpha}(T_{i},P)$ and $\tau_{\beta}(T_{i},P)$ as functions of pressure. When $\tau_{\alpha}(T_{i},P_{g})=100$~s, it can be seen that $\tau_{\beta}(T_{i},P_{g})$ is independent of $T_{i}$.

In most of the cases cited including figure~\ref{NCW_fig02}, the JGX $\beta$-relaxation shows up as a prominent loss peak for all $T$ and $P$ and the peak frequencies $f_{\beta}(T,P)$ are nearly the same, and the invariance of
\[
	\log[\tau_{\alpha}(P,T)]-\log[\tau_{\beta}(P,T)]
\]
at constant $\log[\tau_{\alpha}(P,T)]$ is experimentally verified. This remarkable property of the JGX $\beta$-relaxation implies that it is inseparable from the $\alpha$-relaxation and a necessary precursor. Further considerations of experimental evidences have reached the conclusion that the thermodynamic-scaling or $TV^{\gamma}$-scaling of the $\alpha$-relaxation originates from the primitive relaxation or the JGX $\beta$-relaxation \cite{NCW_bib35}. From these results, one cannot afford to ignore the JGX $\beta$-relaxation in any theory solving the glass transition problem.

\section{Caged dynamics is terminated by the JGX $\beta$-relaxation}
\label{NCW_sctn04}

\begin{figure}[!b]
\centerline{\includegraphics[width=0.6\textwidth]{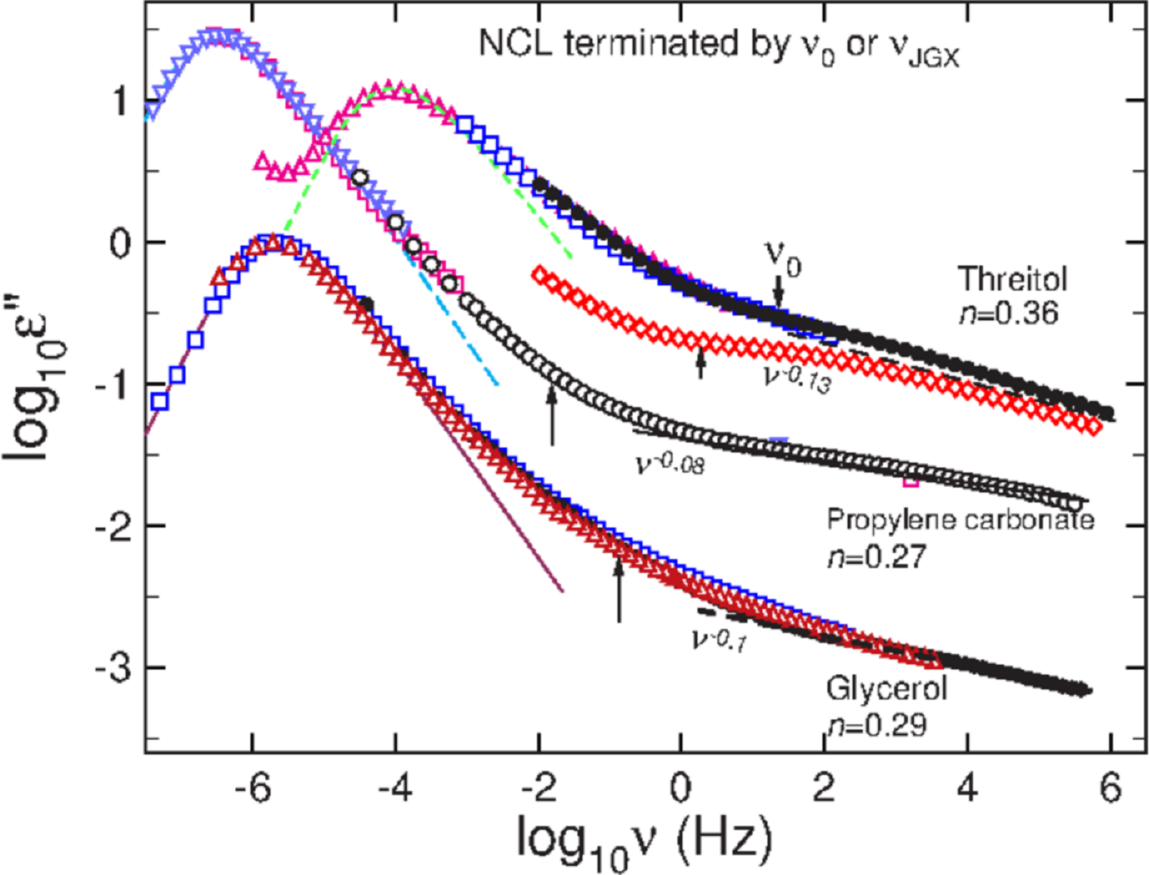}}
\caption{(Colour online) Dielectric loss spectra of glycerol, propylene carbonate, and threitol, to show that the NCL at high frequencies is terminated by the primitive relaxation frequency $\nu_0\approx\nu_{\beta}$. The data of glycerol were scaled down by a factor to avoid an overlap with the data of the two other glass-formers. The lines are the fit of the $\alpha$-loss peak by the Fourier transform of the Kohlrausch function with values of $n$ given in the figure.}
\label{NCW_fig03}
\end{figure}

In the CM, another role of the primitive relaxation is the decay of the cages and the termination of the cage dynamics of glass-formers and ionic conductors as well, and hence from relation~\eqref{NCW_eqn04} that the JGX $\beta$-relaxation plays the same role. This has been demonstrated before \cite{NCW_bib17,NCW_bib24,NCW_bib36} as well as in figure~\ref{NCW_fig03} for three molecular glass-formers, propylene carbonate, glycerol, and threitol.

\begin{figure}[!t]
\centerline{\includegraphics[width=0.75\textwidth]{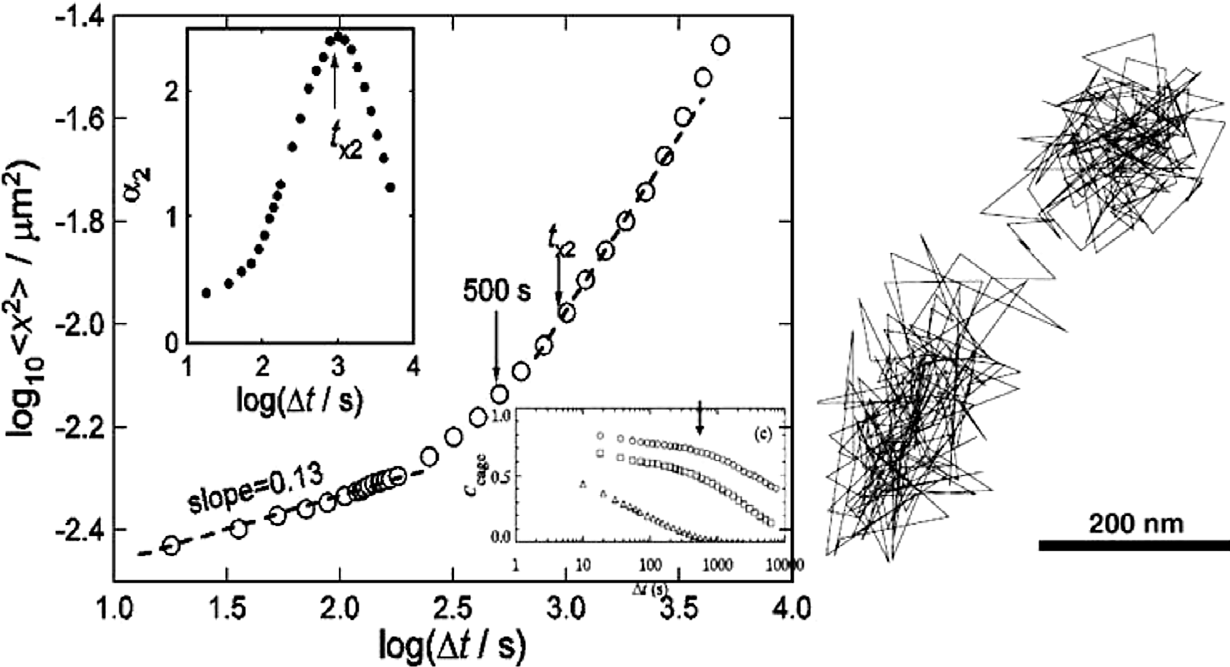}}
\caption{(Left) Mean square displacement $\langle \Delta x^{2}(\Delta t)\rangle$ for volume fractions $\phi=0.56$ from Weeks et al.\ \protect\cite{NCW_bib37}. One vertical arrow indicates $\Delta t=500$~s, the time when a typical particle shifts position and leaves the cage determined by confocal microscopy experiment \protect\cite{NCW_bib37}. The other vertical arrow indicates the time $t_{x2}$ when the non-Gaussian parameter $\alpha_{2}(\Delta t)$ assumes its maximum as shown in the inset on the upper-left corner. The dashed line that has a slope 0.13 indicates the NCL regime. The inset on the lower-right corner is a plot of the cage correlation function $C_{\text{cage}}(\Delta t)$ against $\Delta t$ for three systems with $\phi=0.56$, 0.52, and 0.46 (from top to bottom) \protect\cite{NCW_bib38}, and the vertical arrow indicates $\Delta t=500$~s. 
(Right) A 2D representation of a typical trajectory in 3D for 100 minutes for $\phi=0.56$ from Weeks et al. \protect\cite{NCW_bib37} to illustrate that particles spent most of their time confined in cages formed by their neighbours and moved significant distances only during quick rare cage rearrangements. The particle shown took $\sim500$~s to shift position. Reproduced from references \protect\cite{NCW_bib37,NCW_bib38} by permission.}
\label{NCW_fig04}
\end{figure}

The same is shown in figure~\ref{NCW_fig04} by the collection of data of colloidal particles suspension at volume fraction $\phi=0.56$ by confocal microscopy by Weeks et al. \cite{NCW_bib37}. The right-hand panel of the figure shows the trajectories of particles that they spent most of their time confined in cages formed by their neighbours and moved significant distances only during quick and rare cage rearrangements. A typical particle takes average time $\tau_{c}$ of 500~s to shift position and leaves the cage. This time, identifiable with the primitive relaxation of the CM is indicated by the vertical arrow in the log-log plot of the MSD, $\langle \Delta x^{2}(\Delta t)\rangle$, against $\Delta t$ in the left-hand panel. It can be seen that $\langle \Delta x^{2}(\Delta t)\rangle \sim (\Delta t)^{c}$ with $c\approx 0.13$ at times shorter than 200~s, during which effectively the particles are all confined in cages because $\tau_{c}$ of 500~s has not yet been reached. The $\langle \Delta x^{2}(\Delta t)\rangle$ for $\Delta t < 200$~s corresponds to susceptibility with frequency dependence $\chi''(\nu)\sim\nu^{-c}$ with $c\approx 0.13$, which is the NCL in the caged regime. The lower-right-hand inset is a plot of the cage correlation function $C_{\text{cage}}(\Delta t)$ against $\Delta t$ for three systems with $\phi=0.56$, 0.52, and 0.46 (from top to bottom) \cite{NCW_bib38}, and the vertical arrow indicates $\Delta t=500$~s. It can be seen that for $\phi=0.56$, $C_{\text{cage}}(\Delta t)$ undergoes a significant decay starting at $\Delta t=500$~s.

The phenomenon was found in ionic conductors in experiments and in simulations, and the NCL is terminated by the primitive conductivity relaxation time \cite{NCW_bib24,NCW_bib39}.

\section{Caged dynamics senses the primary ($\alpha$) and the secondary (JGX $\beta$) glass transitions}

In figure~\ref{NCW_fig01} one can see that the temperature dependence of $\langle u^{2}(T)\rangle$ changes from weaker to stronger on crossing $T_{\text{g}}$ from below although data at higher temperatures are not shown. More data showing this property can be found in the review \cite{NCW_bib06}. The $\langle u^{2}(T)\rangle$ data representing the caged dynamics are accessed by neutron scattering measurements of $\langle u^{2}(T)\rangle$ at short times less than 1~ns, while glass transition occurs when $\tau_{\alpha}(T)$ reaches a long time $\sim 10^{3}$~s. There is a disparity of more than 12 orders of magnitude in time scales of $\langle u^{2}(T_{\text{g}})\rangle$ and $\tau_{\alpha}(T_{\text{g}})$ and yet it is the relation. The fact that it is a remarkable property was appreciated by Angell \cite{NCW_bib40}, and others. Obviously this property reflects that caged dynamics is sensitive to density, but the question is why? The answer comes from the coupling of the caged dynamics to the JGX $\beta$-relaxation, which follows from the former and is terminated by the latter as shown in section~\ref{NCW_sctn04}, and also from section~\ref{NCW_sctn03} that the JGX $\beta$-relaxation time $\tau_{\beta}(T)$ is density dependent.

A neutron scattering study of 1,4-PB by Frick et al.\ \cite{NCW_bib41} found that the static structure factor $S(Q)$ is the same at different ($T,P)$ combinations while maintaining the density constant, but the $\alpha$-relaxation and the fast process both change. The fast relaxation shows up as a flat susceptibility minimum, and thus it is the nearly constant loss,
$\chi''(\omega)=A(T)\omega^{-c}$
with 0$<c\ll$1, of caged dynamics. For two ($T,P$) combinations, the one with a higher temperature has shorter $\alpha$-relaxation time and higher intensity $A(T)$ of the nearly constant loss. This correlation is another evidence of the connection of caged dynamics to the merged $\alpha$ and the JGX $\beta$ relaxations. It was also explained in reference \cite{NCW_bib42}.

\begin{figure}[!h]
\centerline{\includegraphics[width=0.45\textwidth]{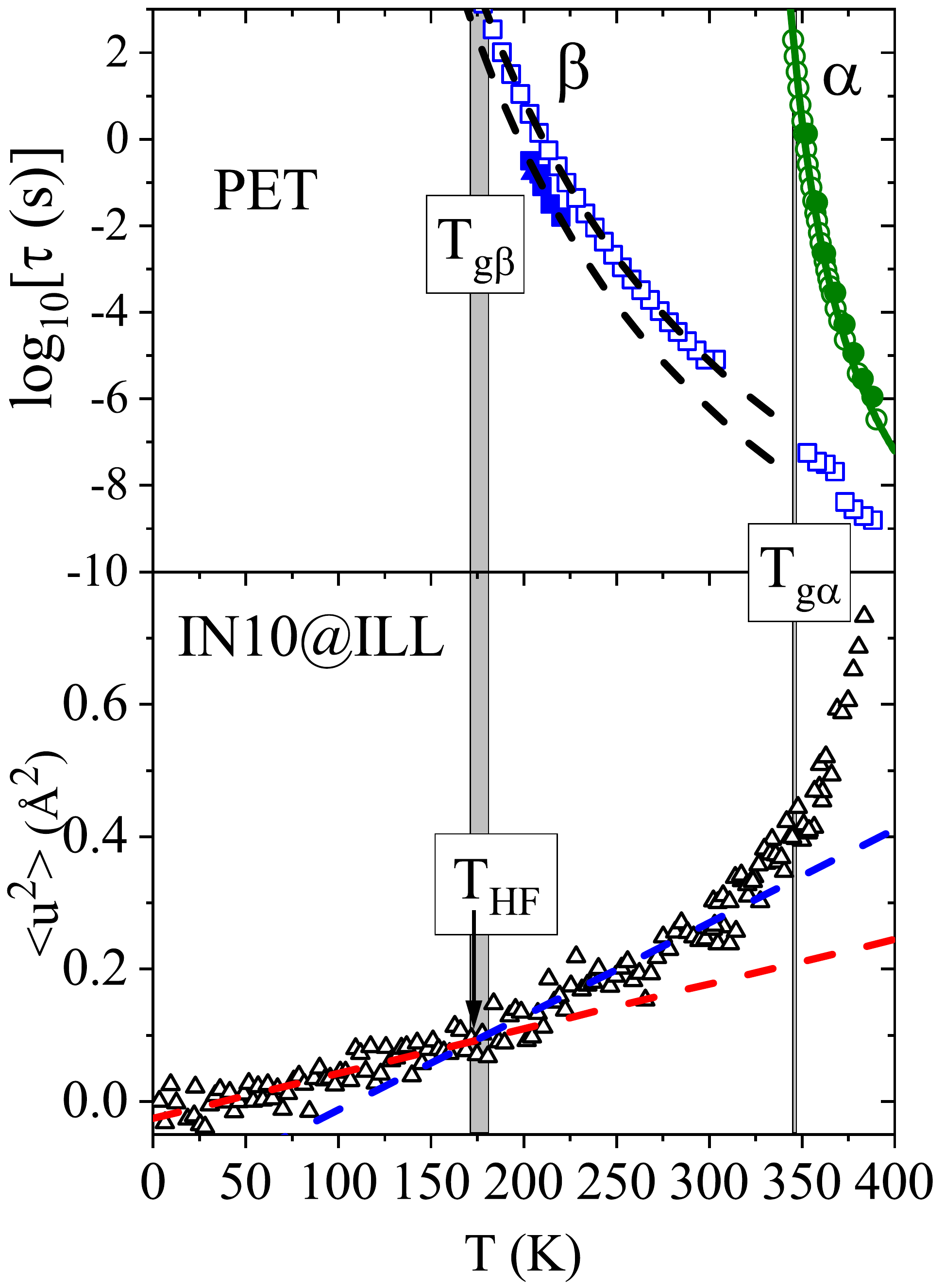}}
\caption{(Colour online) Upper panel: logarithm of the $\alpha$- (green closed) \protect\cite{NCW_bib52} and (open circles) \protect\cite{NCW_bib53} and JGX $\beta$-relaxation time (blue open and closed squares, from dielectric \protect\cite{NCW_bib52,NCW_bib53} and dynamic mechanical spectroscopy data \protect\cite{NCW_bib54}, respectively) of PET plotted as a function of temperature. The low temperature gray shaded area is bracketed from the $T_{\text{g}\beta}$ values obtained when $\tau(T_{\text{g}\beta})=10^{3}$~s for dielectric and dynamic mechanical data. The narrow high temperature shaded area marks the interval for the values of $T_{\text{g}\alpha}$ obtained from the two sets of data from references \protect\cite{NCW_bib52} and \protect\cite{NCW_bib53}. The solid curve is a Vogel-Fulcher-Tammann fit to the $\alpha$-relaxation. The dashed lines are Arrhenius fits for the $\beta$-relaxation times. 
Lower panel: mean square displacement, $\langle u^{2}\rangle$, of amorphous PET (open black triangles) as measured at the IN10 spectrometer plotted as a function of temperature. Data are from reference \protect\cite{NCW_bib55} and redrawn. Data above 383~K are not shown, being potentially affected by cold crystallization. The two dashed lines are linear regressions optimized in the low and intermediate temperature region. The arrow indicates the location of $T_{\text{HF}}$ as obtained from the crossing of the two linear regressions.}
\label{NCW_fig05}
\end{figure}

The paper \cite{NCW_bib43} used terahertz time-domain spectroscopy (THz-TDS) to study the dynamics of the polyalcohols, glycerol, threitol, xylitol, and sorbitol, at temperatures from below to above the glass transition temperature $T_{\text{g}}$. On heating the glasses, they observed that the dielectric losses, $\varepsilon''(\nu)$ at $\nu=1$~THz, increase monotonously with temperature and change the dependence at two temperatures, one at the $T_{\text{g}}$ and another one at a lower temperature $T_{\text{HF}}$. The observed $\varepsilon''(\nu)$ or $\chi''(\nu)$ is the nearly constant loss associated with caged dynamics. Before the publication of the paper by Sibik et al. we already knew that the JGX $\beta$-relaxation is density dependent. Hence, it gives rise to the JGX $\beta$-glass transition at $T_{\text{g}\beta}$ where $\tau_{\beta}(T_{\text{g}\beta})=10^{3}$~s, and temperature dependence of density changes on crossing $T_{\text{g}\beta}$. Also we knew that caged dynamics is connected to the JGX $\beta$-relaxation. Based on this knowledge, we interpreted the $T_{\text{HF}}$ of the polyalcohols from Sibik et al.\ as $T_{\text{g}\beta}$, and confirmed this interpretation by determining $T_{\text{g}\beta}$ by extrapolating the Arrhenius $T$-dependence of $\tau_{\beta}(T)$ to 10$^{3}$~s. The initial success led us to look into the literature of experimental studies in the fast caged dynamics in the glassy state of various glass-formers at high frequencies. The appropriate experimental techniques include neutron scattering, Brillouin scattering, $^{2}$H-NMR spin-lattice relaxation, and high frequency dielectric spectroscopy. In several papers we reported our analyses of the published data to show in many different glass-formers that the intensity of the caged dynamics exhibits a two-step increase at $T_{\text{g}\beta}$ and $T_{\text{g}}$. The phenomenon is general and found in polyalcohols \cite{NCW_bib44}, pharmaceuticals \cite{NCW_bib44}, many amorphous polymers \cite{NCW_bib45}, several small molecular van der Waals glass-formers \cite{NCW_bib46,NCW_bib47}, metallic glass \cite{NCW_bib48}, carbohydrates \cite{NCW_bib49}, and proteins \cite{NCW_bib50,NCW_bib51}. In some cases, $T_{\text{g}\beta}$ was determined directly by calorimetry, positronium annihilation lifetime spectroscopy (PALS). In figure~\ref{NCW_fig05} we show  an example from poly(ethylene terephthalate) of the step change of caged dynamics intensity at $T_{\text{g}\beta}$ \cite{NCW_bib45}.

\section{Summary and conclusions}

The abundant experimental data in the past twenty years have shown that the faster processes, including the caged dynamics and the JGX $\beta$-relaxation, are connected to the slower $\alpha$-relaxation. Manifested in various ways in experiments and simulations, the effects from the coupling clearly confirm that the dynamic and thermodynamic properties of the $\alpha$-relaxation are strongly connected with those of the faster processes. Invoking the principle of causality, we can conclude that the faster processes are indispensable precursors of the $\alpha$-relaxation, and the dependence of the terminal $\alpha$-relaxation time on density and entropy actually originates or inherits from that of the faster JGX $\beta$-relaxation time. Therefore, it is absolutely necessary to consider the faster process in solving the glass transition problem. Any effort to tackle exclusively the $\alpha$-relaxation neglects the fundamentals, and is like putting the cart before the horse. The seminal paper by Scopigno, Ruocco, Sette and Monaco in 2003 \cite{NCW_bib01} and other related papers had made this important point amply clear.

\section*{Acknowledgements}

This work was supported by National Basic Research Program of China (973 Program\linebreak 
No.~2015CB856805), National Natural Science Foundation of China (NSFC) (Grant Nos.~51871193J, 11474247) and Key Research and Development Program of Hebei Province (No.~18391502D).

\ukrainianpart

\title{Абсолютно необхідно враховувати динаміку найближчого оточення та JGX $\beta$-релаксацію в 
розв'язанні проблеми переходу в стан скла}
\author{K.Л. Нгаї\refaddr{label1,label2}, С. Капаччіолі\refaddr{label3}, Л.М. Ванг\refaddr{label2}}
\addresses{%
\addr{label1} CNR-IPCF,  I-56127 Піза, Італія
\addr{label2} Державна ключова лабораторія науки та технології метастабільних матеріалів, Університет Яншану,
Кінгхуангдао, Хебей 066004, Китай
\addr{label3} Фізичний факультет, Університет Пізи, I-56127, Піза, Італія
}

\makeukrtitle

\begin{abstract}
\tolerance=3000%
Визначна стаття 2003 року під назвою ``Чи є крихкість рідини захована у властивостях її скла?'' авторства 
Tулліо Скопіньйо, Джанкарло Руокко, Франческо Сетте і Джуліо Монако повідомляла, що властивості структурної
 $\alpha$-релаксації склоформуючих систем є вже присутні у швидшій динаміці найближчого оточення. Це важливе
 відкриття має далекойдучий наслідок для процесів, швидших за структурну $\alpha$-релаксацію, що не 
 можуть бути проігноровані в розв'язанні проблеми переходу в стан скла. Відтоді,
 експерименти та комп'ютерні моделювання, виконані для багатьох склоформуючих систем з різними хімічними та фізичними
 структурами знайшли сильний зв'язок $\alpha$-релаксації з не лише динамікою найближчого оточення, але й 
 з вторинною релаксацією спеціального типу, названою JGX $\beta$-релаксацією. Загальні результати вказують,
 що ці швидкі процеси є невід'ємні від  $\alpha$-релаксації, та будь-яка спроба розв'язати проблему 
 переходу в стан скла має брати до уваги цей факт. Приклади таких зв'язків подані в цій статті, щоб з'ясувати 
 розробки та прогрес, зроблені після надихаючої публікації Скопіньйо зі співавторами.
\keywords перехід в стан скла, скловидні системи, динаміка найближчого оточення, вторинна релаксація 

\end{abstract}

\end{document}